\documentstyle[preprint,pra,aps,epsfig]{revtex}
\begin{document}
\tighten
\def\question#1{{{\marginpar{\tiny \sc #1}}}}
\draft
\title{NewtonPlus: Approximating Relativistic Effects in Supernova Simulations}
\author{Christian Y. Cardall$^{1,2,3}$ and Anthony Mezzacappa$^1$}
\address{$^1$Physics Division, Oak Ridge National Laboratory, Oak Ridge,
 	TN 37831-6354 \\
	$^2$Department of Physics and Astronomy, University of Tennessee,
	Knoxville, TN 37996-1200 \\ 
	$^3$Joint Institute for Heavy Ion Research, Oak Ridge National
	Laboratory, Oak Ridge, TN 37831-6374}
\date{Talk presented at ``Physics Potential of Supernova II Neutrino 
Detection,'' held February 15-16, 2001 in Marina del Rey, California.}
\maketitle

\begin{abstract}
We propose an approximation to full relativity that captures the 
main gravitational effects of dynamical importance in 
supernovae. The
conceptual link between this formalism and the Newtonian limit is
such that it could likely be implemented relatively easily in existing
multidimensional
Newtonian gravitational hydrodynamics codes employing a Poisson solver.
As a test of the formalism's utility, we display results for rapidly rotating
(and therefore highly deformed) neutron stars.
\end{abstract}


\section{Introduction}
\label{sec:intro}

The collapsed cores of massive stars are relativistic bodies.
Order-of-magnitude estimates of the gravitational potential,
infall and outflow velocities, equatorial linear velocity,
and microscopic nucleon velocity---computed using typical values
of mass, radius, and rotation periods of pulsars---suggest that
relativity cannot be neglected. More importantly, detailed comparisons of 
Newtonian and relativistic supernova simulations in spherical symmetry show
more compact cores and higher neutrino luminosities and average energies
in relativistic treatments \cite{lieb00,brue01}. 
Because of indications that small (several percent) 
variations in, for example, neutrino
luminosities and shock stagnation radius 
can make the difference in a successful explosion 
\cite{jank96}, it is clear
that even modest relativistic effects 
comprise an indispensible component of realism in supernova studies.
 
The multidimensional nature of supernovae must also be recognized as
an important aspect of realism. Various observations, especially
data from SN 1987A, point to the asphericity of supernova explosions
(see e.g. \cite{jank96} for an overview, and \cite{wang01} for a
recent polarimetry analysis of several supernovae). 
Convection may play an important role in the explosion mechanism
\cite{epst79,smar81,arne87,beth87}, and the differing results in
various simulations in up to two dimensions (2D)
\cite{wils93,mill93,hera94,burr95,jank96,mezz98a,mezz98b} show
that further study is needed. 
Ultimately consideration of the third dimension will be necessary.
Based on an initial exploration of 3D effects, it has been reported 
\cite{jank96}
that the sizes of convective cells in 3D simulations are about half as large
as in 2D simulations. Moreover, rapid rotation can significantly affect
the strength and spatial distribution of convection \cite{frye00}.
Detailed studies of magnetic field generation, 
jet formation, and neutron star kicks also invite 3D treatments.

Since neutrinos---which carry away about 99\% of the gravitational binding
energy released in the collapse---are believed to drive the explosions,
accurate neutrino transport is also essential to realistic simulations.
Simulations with Boltzmann transport have been recently performed in
spherical symmetry \cite{mezz00,ramp00,lieb00}.

Including all the physics necessary for realism in a supernova simulation
is a daunting task. The multidimensional simulations mentioned above,
which had simplified neutrino transport, taxed the computational resources
of their time; the same is true of the recent simulations involving Boltzmann
transport in spherical symmetry. Adding
general relativity to the list of desired physics makes things all the
more challenging. While numerical relativity has been successful in 
spherical and axisymmetric cases, ``...in the general three-dimensional
(3D) case which is needed for the simulation of realistic astrophysical
systems, it has not been possible to obtain stable and 
accurate evolutions...,'' and it is argued that the difficulties are
more fundamental than insufficient resolution \cite{alcu00}.

In order to overcome the difficulties associated with
3D general relativistic simulations and to
save resources for 3D hydrodynamics and accurate neutrino
transport, an approximate multidimensional 
treatment of gravity that captures the 
phenomena of dynamical importance in supernovae would be desirable.
The list of new gravitational phenomena introduced by relativity 
includes the nonlinearity of the gravitational field, the inclusion
of all forms of energy and stress as sources, and gravitational waves.
The first two of these effects are of dynamical importance in 
supernovae, while gravitational waves will probably not exert a
strong back-reaction (unless bar or breakup instabilities of some sort
become operative in the core). The post-Newtonian expansion is systematic
and useful in perturbative applications, but in the present context, 
an approach that probes the nonlinear nature of gravity more deeply
would be desirable. A method that could be incorporated into existing
Newtonian hydrodynamics codes would be even more useful.
We here describe such an approximation---which we call 
``NewtonPlus''---and present results
for rapidly rotating (and therefore highly deformed) neutron stars, which show
that this simple ``NewtonPlus'' approach to gravity is indeed a significant
improvement over the Newtonian limit. A more detailed exposition is
given in Ref. \cite{card01}.

\section{Einstein Equations in the NewtonPlus Approximation}

Use of the metric
\begin{equation}
   ds^2 = -(1+2\Phi) dt^2+(1-2\Phi) d{\bf x}^2
\label{newtonMetric}
\end{equation} 
in the Einstein equations yields the Newtonian limit, provided the 
gravitational potential $\Phi\ll 1$ and velocities (including microscopic
velocities, large values of which lead to significant stresses) 
are much less than the speed of light. In order to capture the nonlinearity
of gravity, the significance of stresses as gravitational sources, 
and relativistic fluid velocities, we propose the use of the following
metric: 
\begin{equation}
ds^2 = -e^{2\Phi+2\delta} dt^2+e^{-2\Phi}(dr^2 + r^2 d\Omega^2),
\label{newtonPlusMetric}
\end{equation}
where $d\Omega^2\equiv d\theta^2+r^2\sin^2\theta\, d\phi^2$. In comparison
with Eq. (\ref{newtonMetric}), the ``linearized'' metric
functions have been promoted here to full exponentials, and a second
metric function, $\delta$,
has been added. Eq. (\ref{newtonPlusMetric}) will
then reduce to the Newtonian case if $\delta\rightarrow 0$; we shall see
that this is in fact the case under conditions
in which the Newtonian limit is valid. This provides a tight conceptual
link with the Newtonian limit. Since it has two independent metric functions,
this ``NewtonPlus'' metric should also
provide an exact solution in spherical symmetry.

The metric, a symmetric $4\times 4$ matrix, has ten independent components, but
the invariance of relativity under coordinate transformations implies
that in reality there are only six degrees of freedom.
It is apparent that the NewtonPlus metric of 
Eq. (\ref{newtonPlusMetric}) contains only two
of the six degrees of freedom that should be present. This means that
examination of the complete set of Einstein equations should reveal 
inconsistencies, but we have argued that these are not too serious in
the supernova environment \cite{card01}. Here we present highlights
of that more detailed discussion, which involves the (3+1)
formulation of the Einstein equations \cite{arno62,misn73,smar78,york79},
in which spacetime is foliated into spacelike slices labeled by a time
coordinate.

We begin with the Hamiltonian constraint. This yields 
\begin{equation}
\nabla^2\Phi = 4\pi e^{-2\Phi}E + {1\over 2}(\partial\Phi)^2
  - {3\over 2} e^{-4\Phi-2\delta}(\partial_t\Phi)^2.
\label{hamiltonianConstraint}
\end{equation}
In this expression $\nabla^2$ is the usual 3D flat-space
Laplacian. The energy density as 
viewed by an ``Eulerian''
observer (i.e., one whose 4-velocity is orthogonal to the
spacelike slices, having covariant components
$n_\mu = (-\alpha,0,0,0)$, where $\alpha$ is the lapse function)
is denoted by $E\equiv T^{\mu\nu}n_\mu n_\nu$; where $T^{\mu\nu}$ is the
stress-energy tensor. 
For a perfect fluid, $E=\Gamma^2(\rho+p)-p$, where $\Gamma=(1-v^2)^{-1/2}$,
$v$ is the magnitude of the fluid velocity as measured by an Eulerian 
observer, and $\rho$ and $p$ are respectively the total energy density
and pressure in the fluid rest frame. We have employed the notation
\begin{equation}
\partial X\,\partial Y\equiv \partial_r X\,\partial_r Y +{1\over r^2}\,
  \partial_\theta X\,\partial_\theta Y +{1\over r^2 \sin^2\theta}\,
  \partial_\phi X\,\partial_\phi Y.
\end{equation}
As expected from the conceptual
link between the Newtonian and NewtonPlus metrics, 
Eq. (\ref{hamiltonianConstraint}) identifies $\Phi$ as a glorified
gravitational potential. In addition to the rest energy, the source
includes internal and kinetic energies and pressure, all boosted by
nonlinear contributions from $\Phi$ itself.

The momentum constraints relate $\partial_t\Phi$ to the 
fluid momentum ${\bf s}=\Gamma^2(\rho+p){\bf v}$, 
where ${\bf v}$ is the physical
velocity measured by an Eulerian observer.
Because the test calculations reported here are of
stationary configurations,
$\partial_t\Phi=0$, and the momentum constraints don't
come into play. The condition $\partial_t\Phi=0$
can be taken as a first approximation in the supernova environment as well,
but it will fail if high density matter moves at 
relativistic velocities. Relativistic velocities may be achieved by
infalling matter outside the shock at late
times, or in outflowing jets or winds; but these situations involve matter at 
low density in comparison with the core. Convection may occur deep
in the core, where densities are high, but this likely involves
nonrelativistic velocities. This means that  
${\bf s}=\Gamma^2(\rho+p){\bf v}$ will be arguably small enough everywhere
for $\partial_t\Phi$ to be neglected.

Finally we consider the evolution equations of $\gamma_{ij}$ and
$K_{ij}$. For the NewtonPlus metric, explicit calculation shows
that the former turn out to be 
identities leading to no new information. The latter give three
different equations
for the subdominant metric function $\delta$ that are probably inconsistent, 
though it is not obvious (to us) how to prove this rigorously.
It turns out that stresses (e.g., pressure) constitute the primary source
terms in the equations determining $\delta$, confirming the expectation
expressed previously, that $\delta$ should vanish as the Newtonian limit
is approached. The observation that $\delta$ will only
be appreciable at the highest densities, where pressure begins to make
a nontrivial contribution in comparison with energy density, suggests
a reasonable path forward. In typical cases it is expected that
the deepest
portion of the core will be roughly spherical, even if rapid rotation
causes an equatorial bulge of lower density material
(e.g., Fig. 2 of
Ref. \cite{bona93}). 
If this is the case, neglecting the angular derivatives of $\delta$
is justified, removing many of the apparent inconsistencies in the
three equations for $\delta$. 
The remaining
discrepancies have to do with angular derivatives of $\Phi$ and 
particular components of the stress tensor that appear in each equation.
As it happens, if one adds all three equations, these remaining discrepancies
disappear. Hence the equation we shall use to determine $\delta$ is
\begin{equation}
{{\partial }_r}{{\partial }_r}
   \delta +\frac{1}{r}
    {{\partial }_r}\delta =4\pi
      {{e }^{-2 \Phi }}
         ( {{{S^{\theta }}}_{\theta }}
       +{{{S^{\phi }}}_{\phi }}
        )-{{[{{\partial }_r}(\Phi+\delta) ]}^2},\label{delta}
\end{equation}
where $S_{ij}\equiv T^{km} h_{ik} h_{jm}$, and
the spacelike projection tensor is defined by $h_{ij}\equiv g_{ij}+n_i n_j$.
When it is recalled that $S^{\theta }_{\ \theta }=S^{\phi }_{\ \phi }=p$
in spherical symmetry, this is precisely the equation obtained in the
spherical case. The source on the right-hand side is to be angle-averaged
in solving for $\delta$.

It should be straightforward to include the solution of $\Phi$ and,
if desired, $\delta$, in existing multidimensional gravitational 
hydrodynamics codes. 
The solution of $\Phi$ would make use of the Poisson solver normally
used to solve for the Newtonian gravitational potential, the only difference
being that one would have to iterate on 
Eq. (\ref{hamiltonianConstraint}) (with the $\partial_t\Phi$
term dropped) to get a self-consistent $\Phi$.
The simplest approximation would be to simply solve for $\Phi$ in
this manner, and ignore $\delta$ altogether. The next level of
approximation would involve solving Eq. (\ref{delta}) for
$\delta$, but ignoring $\delta$ on the right hand side. Since $\delta$
is already something of a correction, $\delta$ appearing on the right
hand side is essentially a ``correction to the correction.'' If desired,
however, Eq. (\ref{delta}) could be solved as it stands, with
iteration being required. 

The equations of hydrodynamics are obtained from the vanishing 
divergences of the baryon flux vector and stress-energy tensor,
employing the NewtonPlus metric. Our study of these equations
\cite{card01} shows that they can be cast in a form similar to
that used by at least one Newtonian multidimensional PPM 
hydrodynamics code, VH-1. We plan to adapt this code to the
NewtonPlus approach in the immediate future.

\section{Testing NewtonPlus Gravity with Rapidly Rotating Stars}

In this section we present calculations of neutron stars undergoing
rapid uniform rotation in order to assess the strengths and weaknesses
of the NewtonPlus approximation to relativistic gravity. Our
models were computed with a code described in Ref. \cite{card01b}, which
was written to compute the structure of relativistic axisymmetric stars.
We have modified the code to include the ability to perform computations
in the Newtonian and various NewtonPlus limits: with vanishing metric
function $\delta$, with ``linearized'' $\delta$ (i.e. ignoring $\delta$ on
the right hand side of Eq. (\ref{delta})), and ``full'' $\delta$
(solving Eq. (\ref{delta}) as written). All of the NewtonPlus limits
solve a two-dimensional (and stationary) version of the
nonlinear Poisson-type Eq. (\ref{hamiltonianConstraint}) for the
enhanced ``gravitational potential'' $\Phi$. For the results presented here,
the high-density portion of
the equation of state (EOS) is taken from Ref. \cite{pcl95}, and is based on a
field-theoretic description of cold dense matter. 
We also performed calculations
with a polytropic EOS of adiabatic index 2, and found qualitatively similar
results.

Panel (a) of Fig. \ref{pcl} 
exhibits mass vs. radius curves for spherical stars. 
It shows that while the Newtonian limit exhibits no maximum
mass with this EOS,\footnote{No turnover in the mass vs. radius curve
appears in the Newtonian limit, 
up to the high-density boundary of the tabulated EOS. 
A configuration with
central baryon mass density (total energy density) of 
$3.07\times 10^{15}$ g cm$^{-3}$ ($4.65\times 10^{15}$ g cm$^{-3}$)
has a gravitational mass of 15.6 $M_\odot$ and radius 18.9 km in 
the Newtonian limit, while the relativistic configuration with this 
central density has a gravitational mass of 1.68 $M_\odot$.
and a radius of 9.40 km. We remind the reader that the Chandrasekhar mass
phenomenon is a property of stars built on a polytropic equation of state
with adiabatic index equal to 4/3 (suitable for white dwarfs), 
but that stars constructed on
``realistic'' nuclear equations
of state do not necessarily exhibit this behavior. Instead, the upper mass
limit of neutron stars derives from the the general relativistic instability
indicated by the turning point in the mass vs. radius curve.} 
the NewtonPlus approximation does yield a maximum mass. 
Even with vanishing $\delta$, the approximation captures this consequence
of nonlinear gravity. The ``linear $\delta$'' approximation follows the exact
relativistic curve until the most dense configurations are reached. Since
pressure is the main source for $\delta$ (see equation (\ref{delta})),
the large pressures associated with such high densities raise $\delta$ to 
large enough values that it cannot be neglected on the right-hand side of
equation (\ref{delta}). 
As expected, 
the ``full $\delta$'' 
approximation is indistinguishable from the relativistic results
in spherical symmetry, where only two metric functions are needed to
describe the spacetime exactly.

Panels (b)-(f) of Fig. \ref{pcl} show various physical parameters of rapidly
rotating configurations. In order to test the NewtonPlus approximation
in a nonspherical setting, we ask the question: Given a definite number
of baryons rotating at a given uniform angular velocity $\Omega$, 
what do the various treatments of gravity do with those baryons? 
(The fact that baryon number is a conserved quantity makes this an
obvious way to compare different treatments of gravity.)
To 
answer this question we have computed constant baryon mass sequences
beginning at zero rotation (marked by squares) and ending at the
mass shedding limit (marked by stars). The value of baryon mass chosen,
1.8 $M_\odot$, is close to the maximum baryon mass of 1.95 $M_\odot$ 
for the equation of state we employed. 
The quantities plotted, as a function of the (uniform)
stellar angular velocity, are
gravitational mass; equatorial radius; total angular momentum;
eccentricity, defined as $1-r_p/r_e$, where $r_p$ and $r_e$ are
respectively the polar and equatorial coordinate radii; and the
linear equatorial velocity. 

In panels (b)-(f), the efficacy of the NewtonPlus approximations
can be judged by choosing a value of angular velocity and seeing how close
the approximate quantities come to the fully relativistic value.
While the ``full $\delta$'' approximation is indistinguishable from full
relativity in the spherical case, the two curves representing these
treatments deviate from one another with increasing angular velocity.
As expected, the angular velocities at mass shedding
of the NewtonPlus approximations are closer to the relativistic values
than the Newtonian case. The NewtonPlus treatments are quite successful at 
approximating the gravitational mass, radius, and eccentricity, while 
the success of the results for angular momentum and equatorial velocity
is more modest. (It is expected that the mass and radius of the collapsed
core are more important to the supernova explosion mechanism than the
angular momentum.)

\section{Conclusion}

Accurate neutrino transport, 3D hydrodynamics, and
relativity are all essential for realistic supernova simulations.
Given the constraints of current hardware, these cannot all be
treated simultaneously with the detail they deserve. We have therefore
presented an approximation to full relativity 
(or set of related approximations) that captures the most relevant
relativistic effects in the quasispherical supernova environment:
nonlinearity creating a deeper potential well and pressure being a
nontrivial source of gravitation.

This ``NewtonPlus'' approach to gravity has a tight conceptual link
with the Newtonian limit that yields certain advantageous features. 
The gravitational portion of 
multidimensional Newtonian calculations involves
only the solution of the Poisson equation for the gravitational potential
$\Phi$, and taking its gradient to find the gravitational force. 
The basic idea of our NewtonPlus approach is to promote the Newtonian metric 
functions---which are linear in  $\Phi$---to full exponentials. We also add
a second metric function, $\delta$, whose main source is pressure; hence this
metric function vanishes in the Newtonian limit. 
The Einstein
equations yield a nonlinear Poisson-type equation for a (now enhanced)
``gravitational potential,'' whose solution in 3D can be obtained
in a manner similar 
to what is currently done in the Newtonian limit. The inconsistencies in 
the Einstein equations arising from the reduced number of degrees of freedom
turn out to be relegated to the subdominant metric function 
$\delta$; they can be 
removed by ignoring angular variations in $\delta$. This is expected
to be successful in the
supernova context because the region where $\delta$ makes the greatest
difference is where pressure is significant in comparison with
rest mass density. Normally, this is the deepest portion of the 
collapsed core, which is roughly 
spherical even when the outer layers bulge at the equator 
due to rapid rotation.\footnote{Ultra-strong magnetic fields
\cite{bocq95} or differential rotation (e.g., Ref. \cite{baum00})
can give rise to off-center density maxima, making the spherical
correction for stresses via the metric function $\delta$ less useful.
The NewtonPlus approximation with $\delta=0$ could still be employed
in such (probably exceptional) cases, however.}
This strategy---allowing the main contribution to the gravitational field
to be multidimensional and nonlinear, while allowing a spherical correction
for the contribution of stresses---reproduces fairly well 
many of the physical characteristics of rapidly rotating relativistic stars.
Importantly, the hydrodynamics equations obtained from the NewtonPlus metric
are of the same form as those used in a popular Newtonian hydrodynamics
algorithm, providing the expectation that existing Newtonian codes might
be adapted fairly easily to the NewtonPlus approach.

\acknowledgements{We thank M. Liebend\"orfer, O. E. B. Messer, 
and W. R. Hix for 
discussions, and M. Liebend\"orfer for access to an unpublished
manuscript, whose section on relativistic hydrodynamics was 
particularly insightful. 
We also thank M. Prakash for a tabulated high-density 
equation of state.
C.Y.C. is supported by a DoE PECASE award, and A.M.
is supported at Oak Ridge National Laboratory, managed by UT-Battelle,
LLC, for the U.S. Department of Energy under contract DE-AC05-00OR22725.}


\begin{figure}
\caption{Panel (a): Mass vs. radius curves for spherical configurations
  computed with various treatments of gravity. Panels (b)-(f): Various
  physical quantities characterizing uniformly rotating configurations,
  plotted as a function of angular velocity. Each curve represents
  a constant baryon mass sequence computed with the treatments of gravity
  labeled in panel (a).}
\label{pcl}
\end{figure}

\epsfig{file=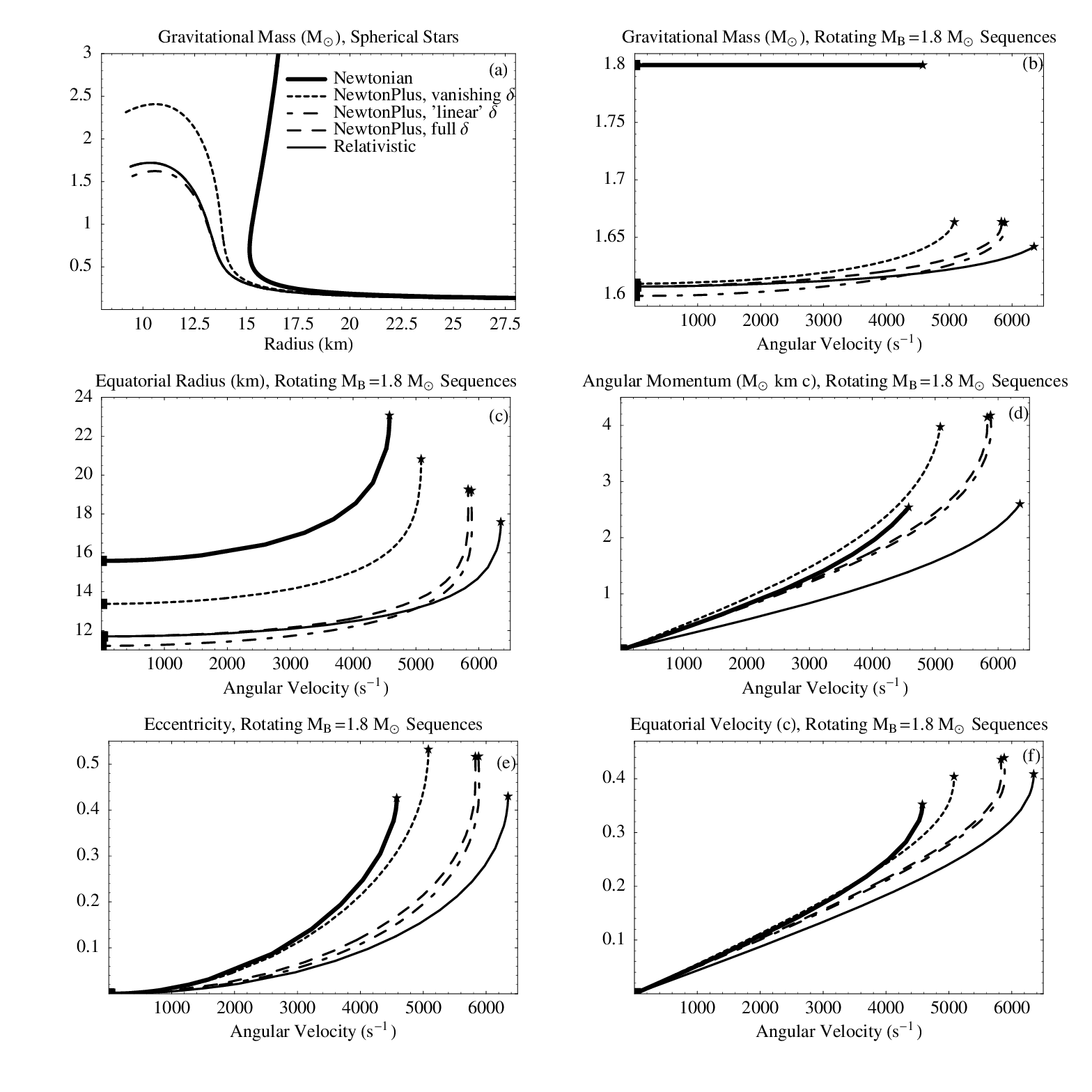,width=6.0in}

\end{document}